\RequirePackage{ifpdf}
\ifpdf 
\documentclass[pdftex]{sigma}
\else
\documentclass{sigma}
\fi

\begin{document}

\allowdisplaybreaks

\renewcommand{\thefootnote}{$\star$}

\renewcommand{\PaperNumber}{016}

\FirstPageHeading

\ShortArticleName{From Noncommutative Sphere to Nonrelativistic Spin}

\ArticleName{From Noncommutative Sphere to Nonrelativistic Spin\footnote{This paper is a contribution to the Proceedings of the XVIIIth International Colloquium on Integrable Systems and Quantum Symmetries (June 18--20, 2009, Prague, Czech Republic).  The full collection is
available at
\href{http://www.emis.de/journals/SIGMA/ISQS2009.html}{http://www.emis.de/journals/SIGMA/ISQS2009.html}}$\!$}

\Author{Alexei A. DERIGLAZOV}

\AuthorNameForHeading{A.A. Deriglazov}

\Address{Dept. de Matematica, ICE, Universidade Federal de Juiz de Fora,
MG, Brazil}
\Email{\href{mailto:alexei.deriglazov@ufjf.edu.br}{alexei.deriglazov@ufjf.edu.br}}

\ArticleDates{Received November 12, 2009, in f\/inal form January 26, 2010;  Published online February 04, 2010}

\Abstract{Reparametrization invariant dynamics on a sphere, being parameterized by angular momentum coordinates, represents an example of noncommutative theory. It can be quantized according to Berezin--Marinov prescription, replacing the coordinates by Pauli matrices. Following the scheme, we present two semiclassical models for description of spin without use of Grassman variables. The f\/irst model implies Pauli equation upon the cano\-ni\-cal quantization. The second model produces nonrelativistic limit of the Dirac equation implying correct value for the electron spin magnetic moment.}


\Classification{81R05; 81R60; 81T75}

\renewcommand{\thefootnote}{\arabic{footnote}}
\setcounter{footnote}{0}

\section{Introduction}

In their pioneer work~\cite{r1}, Berezin and Marinov have suggested
semiclassical description of a spin based on anticommuting variables. Their prescription can be shortly resumed as follows.
For nonrelativistic spin, an appropriate Lagrangian reads
$\frac{1}{2}(\dot x_i)^2+\frac{i}{2}\xi_i\dot\xi_i$, where the spin inner space is constructed from vector like Grassmann variables $\xi_i$, $\xi_i\xi_j=-\xi_j\xi_i$. Since the Lagrangian is linear on $\dot\xi_i$, their conjugate momenta coincide with $\xi$,
$\pi_i=\frac{\partial L}{\partial\dot\xi_i}=i\xi_i$. The relations represent second class constraints of a Hamiltonian formulation and are taken into account by transition from Grassmann Poisson bracket to the Dirac one, the latter reads
\begin{gather}\label{01}
\{\xi_i, \xi_j\}=i\delta_{ij}.
\end{gather}
Dealing with the Dirac bracket, one can resolve the constraints, excluding the momenta from consideration. It gives very economic scheme for description of a spin: there are only three spin variables $\xi_i$. In accordance with equation~(\ref{01}), canonical quantization is performed replacing the variables by Pauli $\sigma$-matrices, $[\sigma_i, \sigma_j]_{+}=2\delta_{ij}$, acting on two dimensional spinor space $\Psi_\alpha$. By this way, formal application of the Dirac method to Grassmann mechanics with constraints allows one to describe both nonrelativistic and relativistic spinning particle on an external electromagnetic f\/ield, see~\cite{r2} for a  review.

It is naturally to ask whether a similar scheme can be realized with
use of commuting variables only (in this relation, let us point out that there is no generalization of Grassmann mechanics on higher spins~\cite{r3}). While description of a spin without Grassmann variables is a problem with a long history~\cite{r4}, search for
spinning particle that would give reasonable classic and quantum theory remains under investigation up to date~\cite{r5}.

To apply the Berezin--Marinov prescription to commuting variables, one needs to construct a dynamical system with an inner space endowed with an algebra that can be realized by $\sigma$-matrices (we discuss the nonrelativistic spin). Due to symmetry properties of a Dirac bracket, it is not possible to arrive at the anticommutator bracket (\ref{01}) working with commuting variables, say $J_i$.
Instead, one can try to produce a bracket of the form $\{J_i, J_j\}=\epsilon_{ijk}J_k$, the latter can also be realized quantum mechanically by $\sigma$-matrices. We discuss such a kind possibility in the present work. Our aim will be to construct a dynamical system that, at the end, admit three degrees of freedom $J_i$ as the spin space basic variables, the latter obey
$SO(3)$ algebra with f\/ixed value of Casimir operator.

One notices that the spin inner space equipped with $SO(3)$-algebra represents an example of noncommutative (NC) system.
Idea of noncommutativity became quite popular after the observation made in~\cite{r6}
that in certain limits string theory can be formulated as an ef\/fective f\/ield theory in NC
space-time. Some well-known physical systems can be also treated from NC point of view, an example is a charged particle conf\/ined to the plane perpendicular to an external magnetic f\/ield. The space becomes noncommutative in the lowest Landau level~\cite{r7}.

Incorporation of NC geometry into the f\/ield theory framework can be naturally achieved using the Dirac method for analysis the constrained theories: NC geometry arises due to the Dirac bracket, after taking into account the constraints presented in the model. It seems to be reasonable approach, since at least the pioneer NC models~\cite{r7, r8, r6} all can be properly treated by this way~\cite{r9, r8}. Moreover, following this way, with any classical mechanical system one associates its NC version~\cite{r10,r11,r12,r13,r14}.

Since NC geometry turns out to be useful tool for reformulation and investigation of some problems in classical and quantum mechanics~\cite{r10,r11,r12,r12,r14,r15,r16} as well as in QFT~\cite{r17}, there are attempts to formulate it on a more fundamental level (see~\cite{r18} for the recent review). One of the barriers here is relativistic (Galilean) invariance. Except a couple of specif\/ic models, compatibility of NC geometry with the symmetries remains an open problem. In $2+1$ dimensions, the Landau problem~\cite{r7} and the Lukierski--Stichel--Zakrzewski higher derivative NC particle~\cite{r8} are compatible with the Galilean invariance. In four dimensions situation is less promising. Similarly to the Snyder NC space~\cite{r19}, $d=4$ NC particle turns out to be compatible with relativistic invariance realized only as a dynamical symmetry~\cite{r10, r20, r21}.

In the present work we avoid the problem since noncommutativity is associated with the inner space, being not only compatible but responsible for appearance the spin representation of $SO(3)$ group.

In this work we concentrate mainly on algebraic construction of the inner spin space, the latter presented in Section~\ref{section2}. Its dynamical realizations are only sketched, technical details will be presented elsewhere. As the dynamical realizations we discuss two dif\/ferent $D=3+1$ spinning particles in Section~\ref{section3}. The f\/irst model implies Pauli equation upon the canonical quantization. The second model produces nonrelativistic limit of the Dirac equation thus leading to correct value for the electron spin magnetic moment.

\section{Noncommutative sphere algebra}\label{section2}

Consider canonical pairs $v_i$, $\pi_j$, $i, j=1, 2, 3$ with the Poisson bracket algebra being
\begin{gather*}
\{v_i, \pi_j\}=\delta_{ij}.
\end{gather*}
To arrive at the desired $SO(3)$ algebra,
we restrict the initial system to lie on some $d=2$ surface of the six dimensional phase space. It will be made in two steps.
First we constrain the coordinates to lie on $d=4$ surface specif\/ied by
\begin{gather}\label{2}
v^2=a^2, \qquad v\pi=0, \qquad a=\mbox{const}.
\end{gather}
The constraints form a second class system, $\{v^2-a^2, v\pi\}=2v^2$. So one takes them into account by transition from the Poisson to Dirac bracket, $\{\;, \;\}_{\rm D1}$, the latter reads
\begin{gather}\label{3}
\{A, B\}_{\rm D1}=\{A, B\}-\{A, v\pi\}\frac{1}{2v^2}\big\{v^2, B\big\}
+\big\{A, v^2\big\}\frac{1}{2v^2}\{v\pi, B\}.
\end{gather}
Here $A$, $B$ are phase space functions. For the phase space coordinates it implies the algebra\footnote{Equation~(\ref{5}) is analogous to the Snyder noncommutative algebra. I am grateful to the referee for this observation.}
\begin{gather}
\{v_i, v_j\}_{\rm D1}=0, \qquad \{v_i, \pi_j\}_{\rm D1}=\delta_{ij}-\frac{1}{v^2}v_iv_j,\nonumber
\\
\label{5}
\{\pi_i, \pi_j\}_{\rm D1}=-\frac{1}{v^2}(v_i\pi_j-v_j\pi_i).
\end{gather}
Constraints are consistent with the Dirac algebra, that is $\{A, v^2-a^2\}_{\rm D1}=0$, $\{A, v\pi\}_{\rm D1}=0$ for any phase space function $A(v, \pi)$. So one can resolve the constraints, keeping four independent coordinates and the corresponding algebra.

We introduce coordinates that turn out to be convenient for canonical quantization of the system. Consider the quantities
\begin{gather}\label{5.1}
J_i\equiv\epsilon_{ijk}v_j\pi_k, \qquad \tilde\pi_1=\pi_1, \qquad \tilde\pi_2=\pi_2, \qquad s=v_i\pi_i,
\end{gather}
where $\epsilon_{ijk}$ is three dimensional Levi-Civita tensor, $\epsilon_{123}=1$, $\epsilon_{ijk}= \epsilon_{[ijk]}$. The quantities $J_i$ obey~$SO(3)$ angular momentum algebra with respect to both Poisson and Dirac brackets.
Equations~(\ref{5.1}) can be inverted
\begin{gather}\label{5.2}
\pi_1=\tilde\pi_1, \qquad \pi_2=\tilde\pi_2, \qquad
\pi_3=-\frac{J_1\tilde\pi_1+J_2\tilde\pi_2}{J_3},
\\ 
v_i=\frac{1}{\pi^2}(\epsilon_{ijk}\pi_jJ_k+s\pi_i),\nonumber
\end{gather}
where $\pi_i$ in the last equality are given by (\ref{5.2}). Hence $J_i$, $\tilde\pi_1$, $\tilde\pi_2$, $s$ can be used as  coordinates of the six dimensional space instead of $v_i$, $\pi_i$. Equations of the surface (\ref{2}) in these coordinates acquire the form $J^2=a^2\tilde\pi^2$, $s=0$.

One can compute the Dirac bracket (\ref{3}) for the new coordinates\footnote{Equivalently, one can start with Poisson bracket of the new variables and construct the Dirac bracket corresponding the constraints $J^2=a^2\tilde\pi^2$, $s=0$.}. To keep the manifest~$SO(3)$ covariance of the formalism, the equation~(\ref{5.2}) prompts to introduce $\tilde\pi_3\equiv-\frac{J_1\tilde\pi_1+J_2\tilde\pi_2}{J_3}$ (or, equiva\-lently, $J\tilde\pi=0$). Then the Dirac brackets of~$J_i$, $\tilde\pi_i$ are
\begin{gather}\label{11}
\{J_i, J_j\}_{\rm D1}=\epsilon_{ijk}J_k,
\\
\label{12}
\{\tilde\pi_i, \tilde\pi_j\}_{\rm D1}=-\frac{1}{a^2}\epsilon_{ijk}J_k, \qquad \{J_i, \tilde\pi_j\}_{\rm D1}=\epsilon_{ijk}\tilde\pi_k,
\end{gather}
while the Dirac bracket of $s$ with any other coordinate vanishes. Since $s=0$ on the constraint surface, it can be omitted from consideration. In resume, the surface (\ref{2}) is now described by coordinates $J_i$, $\tilde\pi_i$ that are constrained by
\begin{gather}
J\tilde\pi=0,
\qquad
J^2=a^2\tilde\pi^2, \label{10}
\end{gather}
and obey the algebra (\ref{11}), (\ref{12}). When $a^2=1$, it is just the Lorentz group algebra written in terms of the rotation $J$ and the Lorentz boost $\tilde\pi$ generators. Notice also that Dirac bracket of the constraints 
(\ref{10}) with any phase space quantity vanishes.

One has already the desired algebra (\ref{11}), but in the system with four independent variables. To improve this, we impose two more second class constraints and construct the corresponding Dirac bracket. To guarantee that the bracket does not modify its form for $J_i$, the equation~(\ref{11}), one of constraints must give vanishing $\rm D1$-bracket with $J_i$. The only possibility is to take (a~function of) Casimir operator of $SO(3)$ algebra. As another constraint, one takes any phase space function that forms second class system with the Casimir operator. The ambiguity in choosing the second constraint suggests that corresponding dynamical realization will be locally invariant theory, see below. Let us take, for example, the constraints
\begin{gather*}
J^2-\frac{3\hbar^2}{4}=0, \qquad \epsilon_{3jk}\tilde\pi_jJ_k=0 \qquad (\mbox{that is} \ v_3=0),
\end{gather*}
Using their bracket, $\{J^2-\frac{3\hbar^2}{4}, \epsilon_{ijk}\tilde\pi_jJ_k\}_{\rm D1}=-2\tilde\pi_3J^2$, one obtains $\rm D2$-Dirac bracket
\begin{gather*}
\{A, B\}_{\rm D2}=\{A, B\}_{\rm D1}
-\big\{A, J^2\big\}_{\rm D1}\frac{1}{2\tilde\pi_3J^2}\{\epsilon_{3jk}\tilde\pi_jJ_k, B\}_{\rm D1}  \nonumber\\
\phantom{\{A, B\}_{\rm D2}=}{}
+\{A, \epsilon_{3jk}\tilde\pi_jJ_k\}_{\rm D1}\frac{1}{2\tilde\pi_3J^2}\big\{J^2, B\big\}_{\rm D1}.
\end{gather*}
In the result  we have $d=2$ surface determined by the constraints
\begin{gather}\label{15}
J\tilde\pi=0, \qquad \tilde\pi^2=\frac{3\hbar^2}{4a^2}, \qquad \tilde\pi_1J_2-\tilde\pi_2J_1=0,
\\
\label{16}
J^2=\frac{3\hbar^2}{4}.
\end{gather}
Equations (\ref{15}) can be used to exclude all $\tilde\pi_i$. Then one deal with the remaining variables $J_i$ obeying $SO(3)$ algebra (\ref{11}) and subject the constraint (\ref{16}).

Finite dimensional irreducible representations of the angular momentum algebra are numbered by the spin~$s$. The condition (\ref{16}) f\/ixes the spin $s=\frac12$. So, $J$ are quantized by
\begin{gather*}
J_i\longrightarrow \hat J_i=\frac{\hbar}{2}\sigma_i,
\end{gather*}
where $\sigma_i$ states for the Pauli matrices. They act on space of two dimensional spinors $\Psi_\alpha$, $\alpha=1, 2$.  As a consequence of the $\sigma$-matrix commutator algebra,
$[\sigma_i, \sigma_j]=2i\epsilon_{ijk}\sigma_k$, the operators $\hat J_i$ obey the quantum counterpart
of the classical algebra (\ref{11})
\begin{gather*}
[\hat J_i, \hat J_j]=i\hbar\epsilon_{ijk}\hat J_k,
\end{gather*}
as well as the constraint (\ref{16}).

\section{Dynamical realizations}\label{section3}

Here we suggest two dif\/ferent spinning particle models that realize the algebraic construction described above.

{\bf Nonrelativistic spinning particle implying the Pauli equation.} Let us start with discussion of spinning part of a Lagrangian. The constraints (\ref{2}) as well as the second constraint from equation (\ref{15}) prompt to write
\begin{gather}\label{19}
L_{\rm spin}=\frac{1}{2g}\dot v^2+g\frac{b^2}{2a^2}+\frac{1}{\phi}\big(v^2-a^2\big), \qquad b^2=\frac{3\hbar^2}{4}.
\end{gather}
Here $g(t)$, $\phi(t)$ are auxiliary degrees of freedom. The corresponding Hamiltonian is given by
\begin{gather*}
H=\frac{g}{2}\left(\pi^2-\frac{b^2}{a^2}\right)-\frac{1}{\phi}\big(v^2-a^2\big)+\lambda_g\pi_g+\lambda_\phi\pi_\phi.
\end{gather*}
where $\pi_g$, $\pi_\phi$ are conjugate momenta for $g$, $\phi$ and $\lambda$ represent lagrangian multipliers for the primary constraints $\pi_g=0$, $\pi_\phi=0$.

The variables $g$, $\phi$ are subject to second class constraints $\pi_{\phi}=0$, $\frac{gb^2}{a^2}+\frac{2a^2}{\phi}=0$ and to the primary f\/irst class constraint $\pi_g=0$. The latter is associated with local symmetry of the action~(\ref{19}) (with a parameter being an arbitrary function $\alpha(t)$)
\begin{gather}\label{22}
\delta v_i=\alpha\dot v_i, \qquad \delta g=(\alpha g)^{.}, \qquad \delta\phi=\alpha\dot\phi-\dot\alpha\phi.
\end{gather}
Imposing the gauge $g=1$, the variables $g$, $\phi$, $\pi_g$, $\pi_{\phi}$ can be omitted from consideration.

Besides one obtains the desirable constraints
\begin{gather}\label{20}
v^2=a^2, \qquad v\pi=0,
\\
\label{21}
\pi^2-\frac{3\hbar^2}{4a^2}=0.
\end{gather}
The pair (\ref{20}) is of second class, while (\ref{21}) represents the f\/irst class constraint\footnote{More exactly, f\/irst class constraint is given by the combination
$\pi^2-\frac{3\hbar^2}{4a^2}+\frac{3\hbar^2}{4a^4}\big(v^2-a^2\big)=0$.}.

Consider now the spinning particle action
\begin{gather*}
S=\int dt\left(\frac{m}{2}\dot x^2+\frac{e}{c}A_i\dot x^i-eA_0
+ \frac{1}{2g}(\dot v_i-\frac{e}{mc}\epsilon _{ijk}v_jB_k)^2+g\frac{b^2}{2a^2}+\frac{1}{\phi}\big(v^2-a^2\big)\right).
\end{gather*}
Here $x_i$, $i=1, 2, 3$, states for spatial coordinates of the particle, and ${\bf B}=\boldsymbol{\nabla}\times{\bf A}$.
Second and third terms represent minimal interaction with the vector potential $A_0$, $A_i$ of an external electromagnetic f\/ield, while the fourth term contains interaction of spin with a magnetic f\/ield. At the end, it produces the Pauli term in quantum mechanical Hamiltonian.

As it has been discussed above, constraints presented in the model allows one to describe it in terms of the variables $x_i$, its conjugate momenta $p_i$, and the spin vector $J_i=\epsilon_{ijk}v_j\pi_k$ (let us point out that in contrast to $v_i$, $\pi_i$, the spin vector $J_i$ turns out to be gauge invariant quantity with respect to the local symmetry (\ref{22})).
One notices that the action leads to reasonable classical theory.
In the variables $x$, $J$, classical dynamics is governed by the Lagrangian equations
\begin{gather}\label{24}
m\ddot x_i=eE_i+\frac{e}{c}\epsilon_{ijk}\dot x_jB_k-\frac{e}{mc}J_k\partial_iB_k,
\\
\label{25}
\dot J_i=\frac{e}{mc}\epsilon_{ijk}J_jB_k.
\end{gather}
It has been denoted ${\bf E}=-\frac{1}{c}\frac{\partial{\bf A}}{\partial t}-\boldsymbol{\nabla}A_0$.
Since $J^2\approx\hbar^2$, the $J$-term disappears from equation~(\ref{24}) in the classical limit $\hbar\rightarrow 0$. Then equation~(\ref{24}) reproduces the classical motion in an external electromagnetic f\/ield. Notice also that in absence of interaction, the spinning particle does not experience an undesirable self acceleration. Equation~(\ref{25}) describes the classical spin precession in an external magnetic f\/ield.

In the Hamiltonian formulation, taking into account the presented constraints, one obtains the only non vanishing Dirac brackets
$\{x_i, p_j\}=\delta_{ij}$, $\{J_i, J_j\}=\epsilon _{ijk}J_k$, and the Hamiltonian
\begin{gather*}
H=\frac{1}{2m}\left(p_i-\frac{e}{c}A_i\right)^2-\frac{e}{mc}J_iB_i+eA_0.
\end{gather*}
Hence canonical quantization of the model implies the Pauli equation
\begin{gather}\label{27}
i\hbar\frac{\partial\Psi}{\partial t}=\left(\frac{1}{2m}\left(\hat{{\bf p}}-\frac{e}{c}{\bf A}\right)^2+eA_0-
\frac{e\hbar}{2mc}\boldsymbol{\sigma}{\bf B}\right)\Psi.
\end{gather}

{\bf Nonrelativistic spinning particle that produces quantum mechanical Hamiltonian corresponding nonrelativistic limit of the Dirac equation.} One remarkable property of the Dirac equation minimally coupled to the vector potential is that it yields the correct gyromagnetic ratio $g=2$ for the electron spin magnetic moment~\cite{r22}. Technically it happens due to the fact that in nonrelativistic limit arises the Hamiltonian $\frac{1}{2m}({\bf p}\boldsymbol{\sigma})^2$ instead of $\frac{1}{2m}({\bf p})^2$. Then the minimal substitution ${\bf p}\rightarrow {\bf p}-\frac{e}{c}{\bf A}$ leads automatically to the Pauli term with correct value of the spin magnetic moment\footnote{Anomalous magnetic moment is discussed in the quantum mechanics framework in~\cite{r23}.} (see the last term in equation (\ref{27})). It is interesting to look for a spinning particle action that would lead to the nonrelativistic limit of the Dirac equation. We propose the action
\begin{gather*}
S=\int dt\left[\frac{3}{2}\left(\frac{\hbar\sqrt m}{2v^2}\epsilon_{ijk}\dot x_i\dot v_jv_k\right)^{\frac{2}{3}}+
\phi\big(v^2-a^2\big)\right].
\end{gather*}
The conventional degree, $\frac{2}{3}$, leads at the end to the desired Hamiltonian. The factor $\hbar\sqrt m$ implies correct dimension of the action. In the Hamiltonian formulation one obtains the constraints
\begin{gather}\label{29}
v\pi=0, \qquad v^2-a^2=0,
\\
\label{30}
vp=0.
\end{gather}
The pair (\ref{29}) is of second class, while (\ref{30}) represents the f\/irst class constraint associated with the
local symmetry
\begin{gather*}
\delta x_i=\alpha v_i.
\end{gather*}
Let us take the gauge\footnote{It is admissible gauge for the motions with $p\pi\ne 0$.} $J^2=\frac{3\hbar^2}{4}$ for the constraint (\ref{30}). On the constraint surface the classical Hamiltonian is given by
\begin{gather*}
H=-\frac{2}{m\hbar^2}({\bf pJ})^2.
\end{gather*}
Canonical quantization ${\bf p}\rightarrow\hat{{\bf p}}=-i\hbar\boldsymbol{\nabla}$, ${\bf J}\rightarrow\hat{{\bf J}}=\frac{\hbar}{2}\boldsymbol{\sigma}$ leads to nonrelativistic limit of the Dirac equation
\begin{gather*}
H=\frac{1}{2m}(\hat{{\bf p}}\boldsymbol{\sigma})^2.
\end{gather*}
It implies correct value of the spin magnetic moment after the minimal substitution
$\hat{{\bf p}}\rightarrow \hat{{\bf p}}-\frac{e}{c}{\bf A}$.

\subsection*{Acknowledgments}
The work has been supported by the Brazilian foundations CNPq (Conselho Nacional de Desenvolvimento
Cient\'{\i}f\/ico  e Tecnol\'ogico - Brasil) and FAPEMIG.

\pdfbookmark[1]{References}{ref}
\LastPageEnding

\end{document}